\documentclass[12pt]{article}
\usepackage{amsmath,epsfig,amssymb,bbm,cite}

\renewcommand{\Im}{{\rm Im}}

\begin{document}

\title{\vbox{
\baselineskip 14pt
\hfill \hbox{\normalsize KUNS-2223}\\
\hfill \hbox{\normalsize YITP-09-50} } \vskip 2cm
\bf $E_{6,7,8}$ Magnetized Extra Dimensional Models 
 \vskip 0.5cm
}
\author{
Kang-Sin~Choi$^{1}$, \
Tatsuo~Kobayashi$^{1}$, \ 
Ryosuke~Maruyama$^{1}$, \\
Masaki~Murata$^{2}$, \ 
Yuichiro~Nakai$^{2}$, \ 
Hiroshi~Ohki$^{1,2}$,  \ Manabu~Sakai$^{2}$\\*[20pt]
$^1${\it \normalsize
Department of Physics, Kyoto University,
Kyoto 606-8502, Japan} \\
$^2${\it \normalsize 
Yukawa Institute for Theoretical Physics, Kyoto University, 
Kyoto 606-8502, Japan}
}

\date{}

\maketitle
\thispagestyle{empty}

\begin{abstract}
We study 10D super Yang-Mills theory with 
the gauge groups $E_6$, $E_7$ and $E_8$.
We consider the torus/orbifold compacfitication 
with magnetic fluxes and Wilson lines.
They lead to 4D interesting models with three families 
of quarks and leptons, whose profiles in 
extra dimensions are quasi-localized because of 
magnetic fluxes.
\end{abstract}

\newpage

\setcounter{page}{1}

\section{Introduction}

Gauge theories with the gauge groups $E_6$, $E_7$ and $E_8$, are 
quite interesting as grand unified theory in particle physics, 
which would lead to 
the standard model at low energy.
All of quarks and leptons are involved in 
the ${\bf 16}$ representations of 
$SO(10)$ and such a ${\bf 16}$ representation 
appears from the adjoint representation and 
${\bf 27}$ representation of $E_6$.
Furthermore, these representations are included 
in adjoint representations of $E_7$ and $E_8$. 
These exceptional gauge theories can be derived 
in heterotic string theory, type IIB string theory with 
non-perturbative effects and F-theory.
Indeed, interesting models have been studied e.g. in
heterotic orbifold models \cite{Kobayashi:2004ud,Forste:2004ie,
Buchmuller:2005jr,Kim:2006hw,Lebedev:2006kn} 
and F-theory~\cite{Beasley:2008dc,Donagi:2008ca,Font:2008id,
Bourjaily:2009vf,Hayashi:2009ge,Marsano:2009ym,Blumenhagen:2009up}.

Recently, extra dimensional field theories play 
an important role in particle phenomenology 
as well as cosmology.
It is one of the most important issues how to realize a chiral spectrum 
in 4D effective field theory
when we start with higher dimensional theory.
One way to realize 4D chiral theory is to introduce 
non-trivial gauge backgrounds like magnetic fluxes.
Indeed, several models with magnetic flux backgrounds have been studied 
in extra dimensional field theory and superstring theory~\cite{Manton:1981es,
Witten:1984dg,Bachas:1995ik,BDL,Blumenhagen:2000wh,
Angelantonj:2000hi,CIM,Troost:1999xn,Alfaro:2006is}.
Furthermore, T-dual of magnetized D-brane models are 
intersecting D-brane models and within the latter framework 
a number of interesting models have been constructed\cite{BDL,
Blumenhagen:2000wh,
Angelantonj:2000hi,Aldazabal:2000dg,
Blumenhagen:2000ea,Cvetic:2001tj}\footnote{
See for a review \cite{Blumenhagen:2005mu} and references therein.}.

The number of zero-modes, that is, the generation number, 
is determined by the magnitudes of magnetic fluxes.
Indeed, one of important aspects in magnetized extra dimensional 
fields is that 
one can solve analytically zero-mode equations on the 
torus with magnetic fluxes.
Their zero-mode profiles are non-trivially quasi-localized.
Such a behavior of zero-mode wavefunctions would be 
phenomenologically important.
For example, when zero-mode profiles are quasi-localized 
far away each other, their couplings would be suppressed.
That could explain small Yukawa couplings for light quarks 
and leptons, and other couplings, which must be 
suppressed from the phenomenological viewpoint.
At any rate, since we know zero-mode profiles explicitly, 
we can compute 4D effective theory concretely at least 
at the tree-level. (See for calculations of 4D effective couplings, 
e.g.~\cite{CIM,DiVecchia:2008tm,Antoniadis:2009bg,Abe:2009dr}.)
Orbifolding is another way to realize 4D chiral spectra.
One can also solve zero-mode equations on the orbifold 
with magnetic fluxes and compute 4D effective 
theory~\cite{Abe:2008fi,Abe:2008sx}.

At the perturbative level, type II string theory can 
realize $U(N)$, $SO(N)$ and $Sp(N)$ gauge groups, 
but not exceptional groups, although 
exceptional gauge theories could be realized 
non-perturbatively.
Thus, the former classes of gauge theories 
like $U(N)$ 
have been studied mainly with 
the magnetic flux backgrounds.
At any rate, exceptional groups are quite interesting from 
the bottom-up phenomenological viewpoint.

Our purpose of this paper is to propose 
phenomenological model building from 
extra dimensional $E_{6,7,8}$ gauge theories 
with magnetic fluxes on the torus and orbifold 
backgrounds.
Our starting point is 10D super Yang-Mills theory with 
gauge groups $E_{6,7,8}$.
We compactify extra 6 dimensions on $(T^2)^3$ or 
the orbifold.
Then, we introduce magnetic fluxes in $(T^2)^3$ as well as 
Wilson lines.
These non-trivial gauge backgrounds, i.e. 
magnetic fluxes and Wilson lines, with/without 
orbifolding would lead to interesting particle phenomenology.
Through this type of model building, we show some 
semi-realistic massless spectra and phenomenological interesting aspects.

This paper is organized as follows.
In section 2, we consider 4D effective theory derived from 
the torus/orbifold compactification with 
magnetic flux and Wilson line background.
Most of them are already known results.
(See e.g. \cite{CIM,Abe:2008fi,Abe:2008sx,Abe-2009}.)
However, we reconsider such backgrounds by 
emphasizing phenomenological implications 
of Wilson lines on magnetized torus models.
In section 3, we study the $E_6$ models, 
and in section 4 we study the $E_7$ and $E_8$ models.
Section 5 is devoted to conclusion and discussion.

\section{Magnetized extra dimensions}

Here we study magnetized torus models 
and orbifold models.
We start with 10D super Yang-Mills theory with the gauge 
group $G$.
Its Lagrangian is written as 
\begin{eqnarray}
{\cal L} &=& 
-\frac{1}{4g^2}{\rm Tr}\left( F^{MN}F_{MN}  \right) 
+\frac{i}{2g^2}{\rm Tr}\left(  \bar \lambda \Gamma^M D_M \lambda
\right),
\end{eqnarray}
where $M,N=0,\cdots, 9$.
Here, $\lambda$ denotes gaugino fields, $\Gamma^M$ is the 
gamma matrix for ten-dimensions and 
the covariant derivative $D_M$ is given as 
\begin{eqnarray}
D_M\lambda &=& \partial_M \lambda - i [A_M, \lambda],
\end{eqnarray}
where $A_M$ is the vector field.
Furthermore, the field strength $F_{MN}$ is given by 
\begin{eqnarray}
F_{MN} &=& \partial_M A_N - \partial_N A_M -i[A_M,A_N].
\end{eqnarray}

Although our starting point here is the 10D super Yang-Mills theory, 
the following discussions can be extended to 
other dimensions, e.g. 6D super Yang-Mills theory.

\subsection{$T^6$ model}

We consider the background $R^{3,1}\times (T^2)^3$,
whose coordinates are denoted by
$x_\mu$ $(\mu=0,\cdots, 3)$ for the uncompact space $R^{3,1}$
and $y_m$ $(m=4, \cdots, 9)$ for the compact space $(T^2)^3$.
We often use complex coordinations $z_d$ $(d=1,2,3)$ 
for the $d$-th torus $T^2_d$, e.g. 
$z_1=y_4+\tau_1 y_5$.
Here, $\tau_d$ denote complex structure moduli of the $d$-th 
$T^2_d$, while the area of $T^2_d$ is denoted by ${\cal A}_d$.
The periodicity on $T^2_d$ is written as 
$z_d \sim z_d +1_d$ and $z_d \sim z_d +\tau_d$.

The gaugino fields $\lambda$ and the vector fields $A_\mu$ and $A_m$ 
are decomposed as 
\begin{eqnarray}
\lambda(x,z) &=& \sum_n \chi_n(x) \otimes \psi_n(z), 
\nonumber \\
A_\mu(x,z) &=& \sum_n A_{n,\mu}(x) \otimes \phi_{n,\mu}(z),
 \\
A_m(x,z) &=& \sum_n \varphi_{n,m}(x) \otimes \phi_{n,m}(z).\nonumber
\end{eqnarray}
Here, we concentrate on zero-modes, $\psi_0(z)$ and 
we denote them as $\psi(z)$ by omitting the subscript ``0''.
Furthermore, the internal part $\psi(z)$ is decomposed 
as a product of the $T^2_d$ parts, i.e. 
$\psi_{(d)}(z_d)$.
Each of $\psi_{(d)}(z_d)$ is two-component spinor, 
\begin{eqnarray}
\psi_{(d)}= \left(
\begin{array}{c}
\psi_{+(d)} \\
\psi_{-(d)}
\end{array}
\right),
\end{eqnarray}
and their chirality for the $d$-th part is denoted by $s_d$.
We choose the gamma matrix for $T^2_d$ as
\begin{eqnarray}
\tilde \Gamma^1_{(d)} = \left(
\begin{array}{cc}
0 & 1 \\ 1 & 0 
\end{array}
\right), \qquad \tilde \Gamma^2_{(d)}=\left(
\begin{array}{cc}
0 & -i \\ i & 0 
\end{array}
\right).
\end{eqnarray}

We introduce the magnetic flux along the $U(1)_a$ (Cartan) direction 
of $G$ on $T^2_d$,
\begin{eqnarray}\label{eq:mag-flux}
F  ={\pi i \over \Im \tau_d} m_{(d)}^a \ (dz_d \wedge d \bar z_d), 
\end{eqnarray}
where $m_{(d)}^a$ is an integer~\cite{toron}.
Here, we normalize $U(1)_a$ charges $q^a$ such that 
all  $U(1)_a$ charges are integers and the minimum satisfies 
$|q^a|=1$.
We assume that 4D N=1 supersymmetry (SUSY) is preserved.\footnote{
4D N=1 SUSY  is preserved by choosing proper values of area ${\cal A}_d$ as
well as $\tau_d$\cite{Troost:1999xn,CIM,Antoniadis:2004pp}.}
The above magnetic flux can be obtained from the vector potential,
\begin{eqnarray}
A(z_d) ={\pi m^a_{(d)} \over \Im \tau_d} \Im (\bar z_d \ dz_d).
\end{eqnarray}
This form of the vector potential satisfies the 
following relations,
\begin{eqnarray}
A(z_d+1_d) &=& A(z_d) +{\pi m^a_{(d)} \over \Im \tau_d} \Im (dz_d), 
\nonumber \\
A(z_d+\tau_d ) &=& A(z_d) +{\pi m^a_{(d)} \over \Im \tau_d} 
\Im (\bar \tau_d \ dz_d).
\end{eqnarray}
Furthermore, these can be represented as the following 
gauge transformations,
\begin{eqnarray}
A(z_d+1_d) = A(z_d) + d \chi_1^{(d)}, \qquad 
A(z_d+\tau_d ) = A(z_d) + d \chi_2^{(d)},
\end{eqnarray}
where 
\begin{eqnarray}\label{eq:chi}
\chi_1^{(d)} = {\pi m^a_{(d)} \over \Im \tau_d} \Im (z_d), \qquad 
\chi_2^{(d)} = {\pi m^a_{(d)} \over \Im \tau_d} \Im (\bar \tau_d \ z_d).
\end{eqnarray}
Then, the fermion field $\psi_{(d)} (z_d)$ with the $U(1)_a$ charge $q^a$ must 
satisfy 
\begin{eqnarray}
\psi_{(d)}  (z_d+1_d) = e^{iq^a\chi_1^{(d)}(z_d)} \psi_{(d)} (z_d), \qquad 
\psi_{(d)}  (z_d+\tau_d) = e^{iq^a\chi_2^{(d)}(z_d)} \psi_{(d)} (z_d).
\end{eqnarray}

By the magnetic flux (\ref{eq:mag-flux}) along the $U(1)_a$ direction,  
all of 4D gauge vector fields $A_\mu$, which have $U(1)_a$ charges, 
become massive, that is, the gauge group is broken from 
$G$ to $G'\times U(1)_a$ without reducing its rank,\footnote{For example, when 
$G=SU(N)$, $G'$ would correspond to $SU(N-1)$.} where 
4D gauge fields $A_\mu$ in $G' \times U(1)_a$ have vanishing 
$U(1)_a$ charges and their zero-modes $\phi_\mu(z)$ have 
a flat profile.
Since the magnetic flux has no effect on the unbroken 
gauge sector, 4D N=4 supersymmetry remains in the 
$G'\times U(1)_a$ sector, that is, there are massless 
four adjoint gaugino fields and six adjoint scalar fields.\footnote{
In string terminology, these adjoint scalar fields correspond to 
open string moduli, that is, D-brane position moduli.
How to stabilize these moduli is one of important issues.}

In addition, matter fields appear from  gaugino fields 
corresponding to the broken gauge part, that is, 
they have non-trivial representations under $G'$ and non-vanishing
$U(1)_a$ charges $q^a$.\footnote{For example, when $G=SU(N)$ and 
$G'\times U(1)_a= SU(N-1)\times U(1)_a $, these matter fields 
have $(N-1)$ fundamental representation under $SU(N-1)$ and 
$U(1)_a$ charge $q^a=1$ and their conjugates.}
The Dirac equations for their zero-modes become 
\begin{eqnarray}
& & \left( \bar \partial_{z_d} + 
\frac{\pi  q^am^a_{(d)}}{2\Im (\tau_d)}  z_d \right) 
\psi_{+(d)}(z_d,\bar z_d) =0, \\
& & \left( \partial_{z_d} - 
\frac{\pi  q^am^a_{(d)}}{2\Im (\tau_d)} \bar z_d \right) 
\psi_{-(d)} (z_d,\bar z_d) =0 ,
\end{eqnarray}
for $T^2_d$.
When $q^am^a_{(d)} >0$, the component $\psi_{+(d)}$ has $M=q^am^a_{(d)}$ 
independent zero-modes and their wavefunctions are  written as \cite{CIM}
\begin{eqnarray}
\Theta^{j,M}(z)=N_M e^{i \pi Mz\Im (z)/ \Im (\tau)} \vartheta 
\left[
\begin{array}{c}
j/M \\ 0
\end{array}
\right]
\left( Mz, M \tau \right),
\end{eqnarray}
where 
$N_M$ is a normalization factor, 
$j$ denotes the flavor index, i.e. 
$j=1,\cdots, M$ and 
\begin{eqnarray}
\vartheta \left[ 
\begin{array}{c}
a \\ b
\end{array} \right]
\left( \nu, \mu \right) 
&=& 
\sum_n \exp\left[ \pi i
   (n+a)^2 \mu + 2 \pi i (n+a)(\nu +b)\right],
\nonumber
\end{eqnarray}
that is, the Jacobi theta-function.
Note that $\Theta^{0,M}(z)=\Theta^{M,M}(z)$.
Furthermore, for $q^am^a_{(d)} > 0$, the other component 
$\psi_{-(d)}$ has no zero-modes.
On the other hand, when $q^am^a_{(d)} <0$, 
the component $\psi_{-(d)}$ has 
$|q^am^a_{(d)} |$ independent zero-modes, but the other component 
$\psi_{+(d)}$ has no zero-modes.

As a result, we can realize a chiral spectrum when 
we introduce magnetic fluxes on all of three $T^2_d$.
That is, since the ten-dimensional chirality of gaugino fields is fixed, 
zero-modes for either $q^a > 0$ and  $q^a < 0$ appear 
with a fixed four-dimensional chirality.
For example, when $q^a> 0$ and $m^a_{(d)} > 0$ for all of $d=1,2,3$, 
only the combination $\psi_{+(1)} \psi_{+(2)} \psi_{+(3)} $ 
has zero-modes and the number of their zero-modes is equal to 
$(q^a)^3m^a_{(1)}m^a_{(2)}m^a_{(3)}$.

Now, let us introduce Wilson lines along the $U(1)_b$ direction of 
$G'$.
That breaks further the gauge group $G'$ to $G'' \times
U(1)_b$ without reducing its rank.\footnote{
For example, when $G'=SU(N-1)$, the Wilson line breaks it 
to $SU(N-2)\times U(1)_b$.}
All of the $U(1)_b$-charged fields including 4D vector, spinor and 
scalar fields become massive because of the Wilson line, when 
they are not charged under $U(1)_a$ and their zero-mode profiles 
are flat.
On the other hand, the matter fields with non-trivial profiles 
due to magnetic flux have different behavior. 
For matter fields with $U(1)_a$ charge $q^a$ and $U(1)_b$ charge $q^b$, 
the Dirac equations of the zero-modes are modified 
by the Wilson line background, $C^b_d = C^b_{d,1} + \tau_d C^b_{d,2}$ as 
\begin{eqnarray}\label{eq:zero-mode-WL-b}
& & \left( \bar \partial_{z_d} + \frac{\pi  }{2\Im (\tau_d)} 
(q^am^a_{(d)}z_d+q^bC^b_d) \right) 
\psi_{+(d)}(z_d,\bar z_d) =0, \\
& & \left( \partial_{z_d} - \frac{\pi  }{2\Im (\tau_d)} 
 (q^am^a_{(d)}\bar z_d +q^b\bar C^b_d) \right) 
\psi_{-(d)} (z_d,\bar z_d) =0,
\end{eqnarray}
where $C^b_{d,1}$ and $C^b_{d,2}$ are real parameters.
That is, we can introduce Wilson lines along the $U(1)_b$ 
direction by replacing $\chi_i^{(d)}$ in 
(\ref{eq:chi}) as~\cite{CIM}
\begin{eqnarray}\label{eq:chi-WL}
\chi_1^{(d)} = {\pi  \over \Im \tau_d} 
\Im (m^a_{(d)}z_d +q^bC^b_d/q^a), \qquad 
\chi_2^{(d)} = {\pi  \over \Im \tau_d} 
\Im (\bar \tau_d (m^a_{(d)}z_d + q^bC^b_d/q^a)).
\end{eqnarray}
Because of this Wilson line, the number of 
zero-modes does not change, but their wave functions 
are shifted as 
\begin{eqnarray}\label{eq:WL-shift}
\Theta^{j,M}(z_d) \rightarrow \Theta^{j,M}(z_d+q^bC^b_d/(q^am^a_{(d)})).
\end{eqnarray}
Note that the shift of zero-mode profiles depend on 
$U(1)_b$ charges of matter fields.
Similarly, we can introduce the Wilson line $C^a_d$ 
along the $U(1)_a$ direction.
Then, the zero-mode wavefunctions shift as 
\begin{eqnarray}\label{eq:WL-shift-ab}
\Theta^{j,M}(z_d+q^bC^b_d/(q^am^a_{(d)})) \rightarrow \Theta^{j,M}
(z_d+C^a_d/m^a_{(d)}+q^bC^b_d/(q^am^a_{(d)})).
\end{eqnarray}
However, the shift due to $C^a_d$ is rather universal shift, but 
the shift by $C^b_d$ depends on the charges $q^b$ of matter fields.
Thus, the shift by $C^b_d$ would be much more important than 
one by $C^a_d$, in particular from the phenomenological viewpoint.

\begin{figure}[t] \begin{center}
\includegraphics[height=3cm]{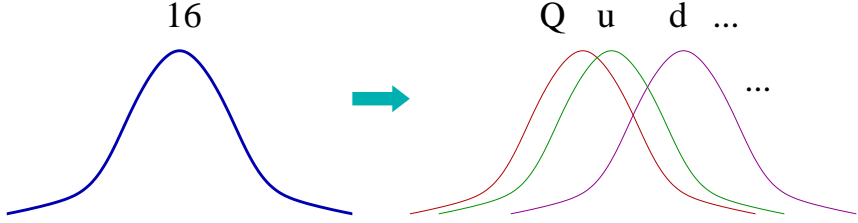}
\caption{Wavefunction splitting by Wilson lines}
\label{fig:WL}
\end{center} \end{figure}

Let us explain more about its phenomenological implications.
Suppose that we introduce magnetic fluxes in a model with 
a larger gauge group $G$ such that they break $G$ to a GUT 
group like $SO(10)$  and this model includes 
three families of matter fields like the ${\bf 16}$ representation, 
corresponding to all of quarks and leptons.
Their 3-point couplings and higher order couplings 
in 4D effective field theory can be computed by 
overlap integral of wavefunctions.
Then, we assume that the $SO(10)$ gauge symmetry is broken 
to $SU(3) \times SU(2) \times U(1)_Y$  by some mechanism.
If zero-mode profiles of quarks and leptons are degenerate 
even after such $SO(10)$ breaking, 
couplings in 4D effective field theory are constrained 
(at the lowest level) by the $SO(10)$ symmetry.
For example, Yukawa matrices have the $SO(10)$ relation, 
that is, Yukawa matrices would be the same between 
the up-sector, the down-sector and the lepton sector.
However, when we break $SO(10)$ to $SU(3) \times SU(2) \times U(1)_Y
\times U(1)$ 
by introducing Wilson lines along the $U(1)_Y \times U(1)$ direction, 
these Wilson lines resolve the degeneracy of zero-mode profiles 
among quarks and leptons.
That is, 
zero-mode profiles of quarks and leptons split depending 
on their $U(1)_Y \times U(1)$ like Figure \ref{fig:WL}.
Then, Yukawa matrices would become different from each other 
among the up-sector, the down-sector and the lepton sector.

Similarly we can analyze 4D massless scalar modes~\cite{CIM}.
We are assuming that 4D N=1 supersymmetry is 
preserved~\cite{Troost:1999xn,CIM}.
Thus, the number of zero-modes and the profiles for 
4D scalar fields are the same as those for their superpartners, 
i.e. the above spinor fields.
For example, for Higgs fields, we study 
zero-modes and their profiles of Higgsino fields.

\subsection{$T^6/Z_2$ model and $T^6/(Z_2\times Z'_2)$ model}

Here, we consider orbifold models.
For simplicity, we start with the $T^2/Z_2$ orbifold.
The $Z_2$ twist acts on the coordinate $z_d$ as 
\begin{eqnarray}
z_d \rightarrow -z_d,
\end{eqnarray}
and the $T^2/Z_2$ orbifold is constructed through 
identifying $z_d \sim -z_d$ on $T^2_d$.
We impose the $Z_2$ boundary condition as 
\begin{eqnarray}\label{eq:z2-bc}
\psi_{\pm (d)}(-z_d) =\pm P \psi_{\pm (d)}(z_d),
\end{eqnarray}
where $P$ is the embedding of the $Z_2$ twist into 
the gauge space.
For non-trivial embedding, the gauge group is broken further.

Magnetic fluxes are invariant under the $Z_2$ twist, 
and we can solve the Dirac equation for zero-modes 
in a way similar to the previous subsection.
Note that 
\begin{eqnarray}
\Theta^{j,M}(-z)=\Theta^{M-j,M}(z),
\end{eqnarray}
where $\Theta^{M,M}(z)=\Theta^{0,M}(z)$.
Thus, we can write the $Z_2$ eigenstates as 
\begin{eqnarray}
\Theta^{j,M}_\pm(z) = \frac{1}{\sqrt 2}
\left(\Theta^{j,M}(z) \pm \Theta^{M-j,M}(z) \right),
\end{eqnarray}
for $j\neq  M/2, M$.
The wavefunctions $\Theta^{j,M}(z)$ for $j=M/2, M$ are 
the $Z_2$ eigenstates with the $Z_2$ even parity.
Either even or odd modes are chosen by the 
boundary condition (\ref{eq:z2-bc}).
Suppose that $q^am^a_{(d)} > 0$.
Then, only the $+$ component $\psi_{+ (d)}$ has zero-modes.
If $P=1$ in Eq.~(\ref{eq:z2-bc}), even zero-modes $\Theta^{j,M}_+(z)$
are chosen.
On the other hand, if $P=-1$, odd zero-modes $\Theta^{j,M}_-(z)$ 
are chosen.
Thus, the number of zero-modes are reduced by orbifolding.
When $M=q^am^a_{(d)}$ is even, the number of even and 
odd zero-modes are equal to $M/2+1$ and $M/2-1$, respectively.
When $M$ is odd, the number of even and 
odd zero-modes are equal to $(M+1)/2$ and $(M-1)/2$, respectively.
These numbers are shown in Table 1.
For example, three families can be obtained from 
even (odd) zero-modes with $M=4$ and 5 (7 and 8).

\begin{table}[t]
\begin{center}
\begin{tabular}{|c||c|c|}\hline
 & $M=$ even & $M=$ odd 
\\ \hline \hline 
even zero-modes & $M/2 +1$ &   $(M+1)/2 $
\\ \hline
odd zero-modes & $M/2 -1$ & $(M-1)/2 $
\\ \hline
\end{tabular}
\end{center}
\caption{The numbers of zero-modes for even and odd wavefunctions.}
\label{even-odd-zero-modes}
\end{table}


Now let us consider the $T^6/Z_2$ orbifold, 
where the $Z_2$ twist acts as 
\begin{eqnarray}\label{eq:z2-10Dtwist}
z_1 \rightarrow -z_1, \qquad z_2 \rightarrow -z_2, \qquad
z_3 \rightarrow z_3.
\end{eqnarray}
Before orbifolding, the gauge sector has 4D N=4 SUSY, but 
this orbifolding reduces it to 4D N=2 SUSY.
For spinor fields, we impose the following $Z_2$ boundary 
condition,
\begin{eqnarray}\label{eq:z2-bc-2}
\psi_{s_1 (1)}(-z_1) \psi_{s_2 (2)}(-z_2) 
\psi_{s_3 (3)}(z_3) =s_1 s_2 P \psi_{s_1 (1)}(z_1) \psi_{s_2 (2)}(z_2) 
\psi_{s_3 (3)}(z_3),
\end{eqnarray}
where $s_d$ denotes the chirality for the $d$-th $T^2_d$, i.e. 
$s_d = \pm 1$.

Furthermore, with the above $Z_2$ twist 
we can consider another $Z_2'$ twist on the 
$T^6/(Z_2 \times Z'_2)$ orbifold.
The $Z_2'$ twist acts as 
\begin{eqnarray}\label{eq:z2-10Dtwist-2}
z_1 \rightarrow -z_1, \qquad z_2 \rightarrow z_2, \qquad
z_3 \rightarrow -z_3.
\end{eqnarray}
Through $Z_2 \times Z'_2$ orbifolding, only 4D N=1 SUSY 
remains even in the gauge sector.
For spinor fields, we impose the following $Z'_2$ boundary 
condition,
\begin{eqnarray}\label{eq:z2-bc-2}
\psi_{s_1 (1)}(-z_1) \psi_{s_2 (2)}(z_2) 
\psi_{s_3 (3)}(-z_3) =s_1 s_3 P' \psi_{s_1 (1)}(z_1) \psi_{s_2 (2)}(z_2) 
\psi_{s_3 (3)}(z_3),
\end{eqnarray}
where $P'$ can be independent of $P$.
Then, depending on the projections $P$ and $P'$, 
even or odd modes for the $d$-th torus remain such as 
$\Theta^{j,M}_{+}(z_d)$ or $\Theta^{j,M}_{-}(z_d)$ and their products 
provide with 
zero-modes on the $T^6/(Z_2 \times Z'_2)$ orbifold.

\section{$E_6$ model}

Here, we consider 10D super Yang-Mills theory with 
the $E_6$ gauge group.

\subsection{$T^6$ model}

We compactify the extra six-dimensions on $T^6$.
We introduce magnetic fluxes (\ref{eq:mag-flux}) 
along the $U(1)_a$ direction, which breaks 
the gauge group, $E_6 \rightarrow SO(10) \times U(1)_a$.
The $E_6$ adjoint representation is decomposed as
\begin{eqnarray}\label{eq:78-rep}
{\bf 78} = {\bf 45}_0 + {\bf 1}_0+ {\bf 16}_1 + \overline {\bf 16}_{-1},
\end{eqnarray}
for $SO(10) \times U(1)_a$.
Here, ${\bf 16}_1$ and $\overline {\bf 16}_{-1}$ correspond to 
the broken part and the corresponding 
gaugino fields appear as matter fields.

For example, we assume magnetic fluxes, 
\begin{eqnarray}\label{eq:mag-3-family} 
m^a_{(1)}=3, \qquad m^a_{(2)}=1, \qquad m^a_{(3)}=1.
\end{eqnarray}
Then, the chiral matter fields corresponding to 
${\bf 16}_1$ and $s_d=(+,+,+)$ have zero-modes, 
but there are no massless modes for $\overline {\bf 16}_{-1}$.
Furthermore, the number of ${\bf 16}_1$ is equal to 
$m^a_{(1)} m^a_{(2)} m^a_{(3)}=3$, that is, 
the model with three families of ${\bf 16}_1$.
Their wavefunctions are written as 
\begin{eqnarray}
\Theta^{j,3}(z_1) \Theta^{1,1}(z_2) \Theta^{1,1}(z_3).
\end{eqnarray}
The flavor structure is determined by the first torus $T^2_1$.
Thus, the massless matter spectrum is realistic, although 
there is no Higgs fields and the gauge sector has 
4D N=4 SUSY.

The $U(1)_a$ symmetry is anomalous.
We assume that its gauge boson become massive by 
the Green-Schwarz mechanism.
Hereafter, we also assume that if other $U(1)$ symmetries 
become anomalous they become massive by the 
Green-Schwarz mechanism.

Here, we break the $SO(10)$ gauge group further 
to the standard model gauge group 
up to $U(1)$ factors, i.e.
$SU(3) \times SU(2) \times U(1)_Y\times U(1)_b$, 
by introducing Wilson lines along $U(1)_Y$ and $U(1)_b$ directions.
The   ${\bf 16}$ representation of $SO(10)$ is decomposed 
under $SU(3) \times SU(2) \times U(1)_Y\times U(1)_b$ as 
\begin{eqnarray}
{\bf 16} = ({\bf 3},{\bf 2})_{1,-1} + (\bar {\bf 3},{\bf 1})_{-4,-1}
+ ({\bf 1},{\bf 1})_{6,-1} + (\bar {\bf 3},{\bf 1})_{2,3}
+ ({\bf 1},{\bf 2})_{-3,3} + ({\bf 1},{\bf 1})_{0,-5}, 
\end{eqnarray}
where we normalize $U(1)_Y$ and $U(1)_b$ charges, such that 
minimum charges satisfy $|q^Y|=1$ and $|q^b|=1$.

By introducing  Wilson lines along $U(1)_Y$ and $U(1)_b$ directions, 
the generation number does not change, but
the zero-mode profiles of three families of 
${\bf 16}$ split differently each other among 
quarks and leptons.
Furthermore, their splitting behaviors depend on 
which torus $T^2_d$ we introduce Wilson lines.
Recall that in this model 
the flavor structure is determined by the first torus $T^2_1$.
For example, when we introduce Wilson lines along $U(1)_Y$ and $U(1)_b$
directions on the second torus $T^2_2$, 
the zero-mode profiles of quarks ($Q,u,d$) and leptons ($L,e,\nu$) 
split as 
\begin{eqnarray}\label{eq:ql-wl-2}
Q&:& \Theta^{j,3}(z^1)\Theta^{1,1}(z^2+C^Y-C^b)\Theta^{1,1}(z^3),
\nonumber \\
u^c&:& \Theta^{j,3}(z^1)\Theta^{1,1}(z^2-4C^Y-C^b)\Theta^{1,1}(z^3), 
\nonumber \\
d^c&:& \Theta^{j,3}(z^1)\Theta^{1,1}(z^2+2C^Y+3C^b)\Theta^{1,1}(z^3), \\
L&:& \Theta^{j,3}(z^1)\Theta^{1,1}(z^2-3C^Y+3C^b)\Theta^{1,1}(z^3),
\nonumber 
\\
e^c&:& \Theta^{j,3}(z^1)\Theta^{1,1}(z^2+6C^Y-C^b)\Theta^{1,1}(z^3), 
\nonumber \\
\nu^c&:& \Theta^{j,3}(z^1)\Theta^{1,1}(z^2-5C^b)\Theta^{1,1}(z^3),
\nonumber  
\end{eqnarray}
where $C^Y$ and $C^b$ are the Wilson lines along 
$U(1)_Y$ and $U(1)_b$ directions.
On the other hand, when we introduce Wilson lines on 
the first torus $T^2_1$, the zero-mode profiles  
of quarks ($Q,u,d$) and leptons ($L,e,\nu$) 
split as 
\begin{eqnarray}\label{eq:ql-wl-1}
Q&:& \Theta^{j,3}(z^1+C^Y/3-C^b/3)\Theta^{1,1}(z^2)\Theta^{1,1}(z^3),
\nonumber \\
u^c&:& \Theta^{j,3}(z^1-4C^Y/3-C^b/3)\Theta^{1,1}(z^2)\Theta^{1,1}(z^3), 
\nonumber \\
d^c&:& \Theta^{j,3}(z^1+2C^Y/3+C^b)\Theta^{1,1}(z^2)\Theta^{1,1}(z^3), \\
L&:& \Theta^{j,3}(z^1-C^Y+C^b)\Theta^{1,1}(z^2)\Theta^{1,1}(z^3),
\nonumber 
\\
e^c&:& \Theta^{j,3}(z^1+2C^Y-C^b/3)\Theta^{1,1}(z^2)\Theta^{1,1}(z^3), 
\nonumber \\
\nu^c&:& \Theta^{j,3}(z^1-5C^b/3)\Theta^{1,1}(z^2)\Theta^{1,1}(z^3).
\nonumber  
\end{eqnarray}
Since the flavor structure is determined by the first torus  $T^2_1$, 
the first case (\ref{eq:ql-wl-2}) preserves the $SO(10)$ flavor 
structure.
However, such flavor structure is deformed in the second case 
(\ref{eq:ql-wl-1}) by Wilson lines.

In order to make this point clear, for the moment 
we assume that our model has 
electro-weak Higgs fields with certain zero-mode profiles, 
although the present model does not include 4D massless 
Higgs fields.
In general, Yukawa couplings are computed by the overlap integral of three 
zero-mode profiles, $\psi_i(z)$, $\psi_j(z)$ and $\psi_k(z)$, 
\begin{eqnarray}\label{eq:yukawa}
y_{ijk} = g \int d^6z \psi_i(z) \psi_j(z) \psi_k(z),
\end{eqnarray}
where $g$ denotes the corresponding coupling in the 
higher dimensional theory.
Such overlap integral for extra 6 dimensions are 
factorized as products of overlap integrals for $T^2$ in 
our case.
Obviously, when we do not introduce Wilson lines along 
$U(1)_Y$ and $U(1)_b$ directions, the zero-mode profiles  
of quarks ($Q,u,d$) and leptons ($L,e,\nu$) do not split 
and the above overlap integral leads to the $SO(10)$ GUT 
relation among quark and lepton Yukawa matrices.
When we introduce the Wilson lines like (\ref{eq:ql-wl-2}), 
the overall factors between up and down quark Yukawa matrices 
as well as lepton Yukawa matrices change, 
but the ratios of Yukawa matrix elements do not change.
On the other hand, in the case with (\ref{eq:ql-wl-1}), 
ratios of Yukawa matrix elements are deformed.

Obviously, other configurations of Wilson lines are possible, 
e.g. $C^Y$ on $T^2_1$ and $C^b$ on $T^2_2$ and so on.
In any case, the flavor structure is determined by which 
Wilson lines we introduce on the first $T^2_1$.
For example, if we introduce only $C^b$ on $T^2_1$, 
the resultant Yukawa matrices would have the $SU(5)$ GUT 
relation.

Thus, the above model is interesting.
Its chiral matter spectrum is realistic and 
the model has the interesting flavor structure, 
although electro-weak Higgs fields do not appear 
and the gauge sector has 4D N=4 SUSY.

\subsection{Orbifold model}

Here, let us study the orbifold model.
First, we introduce the following magnetic fluxes,
\begin{eqnarray}
m^a_{(1)}=4, \qquad m^a_{(2)}=1, \qquad m^a_{(3)}=0,
\end{eqnarray}
in order to break $E_6$  to $SO(10) \times U(1)_a$.
We consider $T^6/(Z_2 \times Z_2')$ orbifold 
with the trivial twists, $P=P'=1$.
Note that all of zero-mode profiles on the third torus are 
flat.
Then, even zero-modes for both first and second tori 
survive through the $Z_2 \times Z_2'$ projection and 
we can realize three families of ${\bf 16}$.
Their wavefunctions for $T^2_1$ and $T^2_2$ are written as 
\begin{eqnarray}\label{eq:wf-orb-3}
\Theta^{j,4}_+(z_1) \Theta^{1,1}(z_2),
\end{eqnarray}
and they have flat profiles for $T^2_3$. 
Note that $\Theta^{1,1}(z_2)=\Theta^{1,1}_+(z_2)$.
The flavor structure is determined by the first torus $T^2_1$.
Furthermore, the $Z_2 \times Z_2'$ twists reduce 
4D N=4 SUSY to 4D N=1 SUSY in the gauge sector.
Hence, the 4D massless spectrum of this orbifold model 
is quite realistic, although electro-weak Higgs fields 
do not appear.

We can construct similar models.
For example, we introduce the magnetic fluxes,
\begin{eqnarray}
m^a_{(1)}=5, \qquad m^a_{(2)}=1, \qquad m^a_{(3)}=0,
\end{eqnarray}
on the $T^6/(Z_2 \times Z_2')$ orbifold.
That also leads to three families of ${\bf 16}$ with 
the wavefunctions $\Theta^{j,5}_+(z_1) \Theta^{1,1}(z_2)$ and 
the flat profile along $T^2_3$.
These wavefunctions, in particular the first $T^2_1$ part, are 
different from (\ref{eq:wf-orb-3}), and 
they lead to different aspects, e.g. in the flavor structure.

Similarly, we can consider the $T^6/Z_2 $ orbifold (\ref{eq:z2-10Dtwist})
with the magnetic fluxes,
\begin{eqnarray}
m^a_{(1)}=4, \qquad m^a_{(2)}=1, \qquad m^a_{(3)}=1,
\end{eqnarray}
and 
\begin{eqnarray}
m^a_{(1)}=5, \qquad m^a_{(2)}=1, \qquad m^a_{(3)}=1.
\end{eqnarray}
Both of them lead to three families of ${\bf 16}$ with 
the wavefunctions,
$\Theta^{j,4}_+(z_1) \Theta^{1,1}(z_2)\Theta^{1,1}(z_3)$
and
$\Theta^{j,5}_+(z_1) \Theta^{1,1}(z_2)\Theta^{1,1}(z_3)$.
Note that the third plane is just $T^2$, but not orbifold.
For the gauge sector, 4D N=2 SUSY remains and 
for $T^2_3$ we can introduce Wilson lines, which we discussed 
in section 2.1.
Moreover, on the $T^6/Z_2 $ orbifold, the following magnetic flux 
\begin{eqnarray}
m^a_{(1)}=1, \qquad m^a_{(2)}=1, \qquad m^a_{(3)}=3,
\end{eqnarray}
also leads to three families of ${\bf 16}$ with 
the wavefunctions 
$\Theta^{1,1}(z_1) \Theta^{1,1}(z_2)\Theta^{j,3}(z_3)$.
Recall that on the $T^6/Z_2 $ orbifold, 
the third torus $T^2_3$ and the others have different 
behaviors from each other.

Now, let us study the possibility for introducing 
electro-weak Higgs fields.
The orbifold has fixed points.
It is possible to put certain modes on such fixed points.
Here, we assume electro-weak Higgs fields as such localized modes, 
although the gauge multiplets and three families are 
originated from 10D bulk modes.
If the gauge symmetry is 
broken to the $SU(3) \times SU(2) \times U(1)_Y\times U(1)_a
\times U(1)_b$ on orbifold fixed points,\footnote{Some 
$U(1)$ factors may be broken.}
we do not need to introduce full multiplets of 
$E_6$, $SO(10)$ or $SU(5)$.
Thus, we could assume only one pair of Higgs doublets on 
orbifold fixed points on some of $T^2_d$, 
and they may be bulk modes on other of $T^2_d$.
The Yukawa couplings among Higgs fields and matter fields 
could be allowed only on fixed points.
Thus, Yukawa couplings are determined by magnitudes of 
zero-mode profiles of quarks and leptons on such a fixed point.
Note that all of matter zero-modes are even functions and 
they have non-vanishing values on fixed points, 
although odd wavefunctions vanish on fixed points.

\section{$E_7$ and $E_8$ models}

\subsection{$T^6$ model}

Similarly, we can study $E_7$ and $E_8$ models.
Their ranks are larger than $E_6$ and their adjoint 
representations include several representations.
The $E_8$ adjoint representation ${\bf 248}$ is decomposed 
under $E_7 \times U(1)_{E8}$ as 
\begin{eqnarray}\label{eq:248-rep}
{\bf 248} = {\bf 133}_0 + {\bf 1}_0 +{\bf 56}_1 + 
{\bf 56}_{-1}+{\bf 1}_2 + {\bf 1}_{-2} .
\end{eqnarray}
Note that we are using $U(1)$ charge normalization such that 
the minimum charge except vanishing charge is equal to one, $|q|=1$. 
Then, the $E_7$ adjoint representation ${\bf 133}$ 
is decomposed under $E_6 \times U(1)_{E7}$ as 
\begin{eqnarray}\label{eq:133-rep}
{\bf 133} = {\bf 78}_0 + {\bf 1}_0 +{\bf 27}_{-2} + 
\overline {\bf 27}_{2},
\end{eqnarray}
 and the ${\bf 56}$ representation of $E_7$ is decomposed 
under $E_6 \times U(1)_{E7}$ as 
\begin{eqnarray}\label{eq:56-rep}
{\bf 56} = {\bf 27}_1 + 
\overline {\bf 27}_{-1} + {\bf 1}_2 + {\bf 1}_{-2}.
\end{eqnarray}
Furthermore, the ${\bf 27}$ representation of $E_6$ 
is decomposed under $SO(10)\times U(1)_{E6}$ as 
\begin{eqnarray}
{\bf 27} ={\bf 16}_1 + {\bf 10}_{-2} + {\bf 1}_4.
\end{eqnarray}
Thus, we can construct various models from 
$E_7$ and $E_8$ models.
Quark and lepton matter fields can be originated from 
several sectors, although such matter fields are originated from 
${\bf 16}$ of the $E_6$ adjoint sector in the models 
of the previous section.
In addition,  the $E_7$ and $E_8$ adjoint representations 
include exotic representations.
Hence, exotic matter fields, in general, appear in 
4D massless spectra.
Instead of $U(1)_{E8} \times U(1)_{E7}$, we use the  
$U(1)_c \times U(1)_d$ basis, such that those charges are related as 
\begin{eqnarray}
q_c=\frac{1}{2}q_{E8} + \frac{1}{2}q_{E7}, \qquad 
q_d=-\frac{1}{2}q_{E8} + \frac{1}{2}q_{E7},
\end{eqnarray}
where $q_c$, $q_d$, $q_{E8}$ and $q_{E7}$ denote 
$U(1)_c$, $ U(1)_d$, $U(1)_{E8}$ and $U(1)_{E7}$ charges, 
respectively.
In addition, we denote $U(1)_{E6}$ by $U(1)_a$ as in section 3.
Also, as in section 3, we use the notation $U(1)_b$, which 
appears through the $SO(10)$ breaking as $SO(10) \rightarrow SU(5)
\times U(1)_b$.

Here, we show just simple illustrating models.
First of all, we can construct almost the same model as 
the $E_6$ models.
For example, we start with the 10D $E_7$ super Yang-Mills theory.
We can introduce magnetic fluxes with the same form in 
$U(1)_{E6}$ as (\ref{eq:mag-3-family}).
Furthermore, we introduce Wilson lines such that the 
gauge group is broken down to 
$SU(3) \times SU(2) \times U(1)_Y$ up to $U(1)$ factors.
Then, we realize three families of quarks and leptons under 
the standard model gauge group, 
that is, the same 4D massless spectrum as one in section 3.1, 
although the gauge sector has partly 4D N=4 SUSY and 
there is no Higgs fields.
Similarly, the same model can be derived from the 10D 
$E_8$ super Yang-Mills theory.
Also the same orbifold models as one in section 3.2 
can be derived from 10D 
$E_7$ and $E_8$ super Yang-Mills theories.

Now, let us consider another illustrating model with different aspects.
We start with the 10D $E_8$ super Yang-Mills theory.
When $E_8$ is broken to the standard model gauge group, there are 
five $U(1)$'s including $U(1)_Y$, i.e., $U(1)_I$ $(I=a,b,c,d,Y)$.
We introduce magnetic fluxes $m^I_{(d)}$ 
along these five $U(1)_I$ directions.
Then,  the sum of magnetic fluxes $M=\sum_I q^I m^I_{(d)}$
appears in the zero-mode Dirac equation for the matter field with 
charges $q^I$.
We require that $\sum_I q^Im^I_{(d)}$  should be integer for 
all of matter fields, that is, the quantization condition of 
magnetic fluxes~\cite{toron}.

For example, five $({\bf 3},{\bf 2})_{1}$ representations 
under $SU(3) \times SU(2) \times U(1)_Y$ 
as well as their conjugates appear from the ${\bf 248}$ adjoint 
representation.
Three of them appear from three ${\bf 27}$ representations 
of ${\bf 248}$, i.e., Eqs.~(\ref{eq:248-rep}), 
(\ref{eq:133-rep}) and (\ref{eq:56-rep}).
In the zero-mode equations of such three $({\bf 3},{\bf 2})_{1}$ matter fields, 
the following sum of magnetic fluxes $\sum_I q^I m^I_{(d)}$ appear
\begin{eqnarray}
m^{Q1}_{(d)} &=& m^c_{(d)} + m^a_{(d)} - m^b_{(d)} + m^{Y}_{(d)}, 
\nonumber \\
m^{Q2}_{(d)} &=& m^d_{(d)} + m^a_{(d)} - m^b_{(d)} + m^{Y}_{(d)},\\
m^{Q3}_{(d)} &=& -m^c_{(d)} -m^d_{(d)} 
+ m^a_{(d)} - m^b_{(d)} + m^{Y}_{(d)}. \nonumber
\end{eqnarray}
In addition, one $({\bf 3},{\bf 2})_{1}$ representation appears from 
${\bf 16}$ of the $E_6$ adjoint ${\bf 78}$ representation 
(\ref{eq:78-rep}) as section 3.
In the zero-mode equation of such $({\bf 3},{\bf 2})_{1}$ matter field, 
the following sum of magnetic fluxes $\sum_I q^I m^I_{(d)}$ appears 
\begin{eqnarray}
m^{Q4}_{(d)} &=& -3 m^a_{(d)} - m^b_{(d)} + m^{Y}_{(d)} .
\end{eqnarray}
Moreover, the $SO(10)$ adjoint ${\bf 45}$ representation also include a 
$({\bf 3},{\bf 2})_{1}$ representation and 
the corresponding matter field has the sum 
of magnetic fluxes $\sum_I q^I m^I_{(d)}$, \footnote{The 
$SO(10)$ adjoint ${\bf 45}$ representation includes another 
$({\bf 3},{\bf 2})$ representation but its $U(1)_Y$ charge is different.}
\begin{eqnarray}
m^{Q5}_{(d)} = 4m^b_{(d)} + m^{Y}_{(d)} ,
\end{eqnarray}
in the zero-mode equation.
Here, we require that all of 
$m^{Q1}_{(d)}$, $m^{Q2}_{(d)}$, $m^{Q3}_{(d)}$, $m^{Q4}_{(d)}$
and $m^{Q5}_{(d)}$ should be integers.
Similarly, we require that $\sum_I q^Im^I_{(d)}$  should be integers for 
all of matter fields with charges $q^I$, which appear from 
the $E_8$ adjoint ${\bf 248}$ representation.
By an explicit computation, it is found that 
the sum $\sum_I q^Im^I_{(d)}$ for any charge 
$q^I$ appearing from ${\bf 248}$ can be written as a linear 
combination of $m^{Q1}_{(d)}$, $m^{Q2}_{(d)}$, $m^{Q3}_{(d)}$, $m^{Q4}_{(d)}$
and $m^{Q5}_{(d)}$ with integer coefficients.
Thus, when all of $m^{Q1}_{(d)}$, $m^{Q2}_{(d)}$, $m^{Q3}_{(d)}$, $m^{Q4}_{(d)}$
and $m^{Q5}_{(d)}$ are integers, 
the sum $\sum_I q^Im^I_{(d)}$ for any charge $q^I$ of 
${\bf 248}$ is always integer.

Using the above notation, 
we introduce the  magnetic fluxes such as ,
\begin{eqnarray}
 & & m^{Q1}_{(1)} = 1, \qquad m^{Q1}_{(2)} = -1, \qquad m^{Q1}_{(3)} = -3, 
\nonumber  \\
 & & m^{Q2}_{(1)} = -1, \qquad m^{Q2}_{(2)} = 0, \qquad m^{Q2}_{(3)} = 1,  
\nonumber \\
 & &  m^{Q3}_{(1)} = -1, \qquad m^{Q3}_{(2)} = 0, \qquad m^{Q3}_{(3)} = 1, \\
 & &  m^{Q4}_{(1)} = -1, \qquad m^{Q4}_{(2)} = 0, \qquad m^{Q4}_{(3)}  = 1,  
\nonumber \\
 & &  m^{Q5}_{(1)} = -2, \qquad m^{Q5}_{(2)} = -1, \qquad m^{Q5}_{(3)}
 = 0. 
\nonumber
\end{eqnarray}
In addition, we also introduce all possible Wilson lines 
on each torus along five $U(1)$ directions.
Then, the gauge group is $SU(3) \times SU(2) \times U(1)_Y$ 
with $U(1)$ factors.

The 4D massless spectrum of this model includes 
the following matter fields under the standard gauge group, 
$SU(3) \times SU(2) \times U(1)_Y$,
\begin{eqnarray}
& & 3 \times \left[ ({\bf 3},{\bf 2})_1 + (\overline {\bf 3},{\bf 1})_{-4} 
+ (\overline {\bf 3},{\bf 1})_{2} + ({\bf 1},{\bf 2})_{-3} 
  + ({\bf 1},{\bf 1})_{6} \right]  \nonumber \\
& & + 8 \left[ ({\bf 1},{\bf 2})_{3}+ ({\bf 1},{\bf 2})_{-3} \right]
\\
& & + 15 \times  \left[  ({\bf 3},{\bf 1})_{4} + 
 (\overline {\bf 3},{\bf 1})_{-4} \right] 
+ 6 \times  \left[  ({\bf 3},{\bf 1})_{-2} + 
 (\overline {\bf 3},{\bf 1})_{2} \right] 
+ 27 \times  \left[ ({\bf 1},{\bf 1})_{6} 
+ ({\bf 1},{\bf 1})_{-6}\right],  \nonumber 
\end{eqnarray}
and $SU(3) \times SU(2)$ singlets with vanishing $U(1)_Y$ charges.
That is, this massless spectrum includes 
three families of quarks and leptons as well as 
eight pairs of up- and down-sectors of electroweak Higgs fields.
In addition, many vector-like matter fields appear, 
but matter fields with exotic representations 
do not appear even in vector-like form.
Such exotic matter fields have (effectively) vanishing 
magnetic flux on one of $T^2_d$.
Then, such fields become massive when we switch on proper Wilson
lines.\footnote{In the limit of vanishing Wilson lines, such 
exotic fields appear in the vector-like form, but 
they become massive for finite values of Wilson lines.}
Thus, this model has semi-realistic massless spectrum, 
although the gauge sector still has 4D N=4 SUSY.
We can write the wavefunctions of these zero-modes.
For example, the zero-mode wavefunctions of 
left-handed quarks are written as 
\begin{eqnarray}
\Theta^{1,1}(z_1+C_1)\Theta^{1,1}(z_2+C_2)\Theta^{j,3}(z_3+C_3/3),
\end{eqnarray}
for $j=1,2,3$,
where $C_d$ denote Wilson lines along five $U(1)$ directions.
Thus, the flavor structure for the left-handed quarks is 
determined by the third torus.
Similarly, we can write zero-mode wavefunctions of the other 
matter fields.
The above massless spectrum includes several vector-like generations of 
right-handed quarks as well as right-handed leptons.
These vector-like generations may gain mass terms.
Thus, the flavor structure of chiral right-handed quarks depends 
on mass matrices of vector-like generations.

Similarly, various models can be constructed within 
the framework of $E_7$ and $E_8$ models with 
magnetic flux and Wilson line backgrounds.
This type of model building would lead to quite interesting models.
We would study these types of model building 
systematically elsewhere.

\subsection{Orbifold model}

The orbifold background with 
$E_7$ and $E_8$ gauge groups can also be studied.
As the $E_6$ model, we can consider the orbifold models 
with the trivial twist $P=1$.
However, here we study the orbifold models with 
non-trivial twists and Wilson lines, which 
break the gauge groups with reducing their ranks, 
in order to show the variety of model building 
on the backgrounds with magnetic fluxes, Wilson lines 
and orbifolding.

As an illustrating model, we start with 
the 10D $E_7$ super Yang-Mills theory on the 
$T^6/Z_2$ orbifold of section 2.2.
The $E_7$ adjoint representation ${\bf 133}$ is 
decomposed under $SO(10) \times SU(2) \times U(1)$ as 
\begin{eqnarray}
{\bf 133} = ({\bf 45},{\bf 1})_0 + ({\bf 1},{\bf 3})_0 +({\bf 1},{\bf 1})_0 +
({\bf 16},{\bf 2})_1 +(\overline{\bf 16},{\bf 2})_{-1}
+({\bf 10},{\bf 1})_{2} +({\bf 10},{\bf 1})_{-2}.
\end{eqnarray}
For example, we assume the following magnetic fluxes 
along this $U(1)$ direction,
\begin{eqnarray} 
m_{(1)}=3, \qquad m_{(2)}=1, \qquad m_{(3)}=1.
\end{eqnarray}
Then, three $({\bf 16},{\bf 2})_1$ zero-modes as well as 
24 $({\bf 10},{\bf 1})_{2}$ zero-modes would appear without 
orbifolding.
Now, let us study non-trivial orbifold twists and Wilson lines, 
e.g. in the $SU(2)$ part.
We concentrate on $({\bf 16},{\bf 2})_1$ matter fields, because 
only these matter fields have non-trivial representations 
under the $SU(2)$ group, that is, the doublet,
\begin{eqnarray}
\left(
\begin{array}{c}
\lambda_{1/2} \\ \lambda_{-1/2} 
\end{array}
\right).
\end{eqnarray}
Before orbifolding, two components, $\lambda_{1/2}$ and 
$\lambda_{-1/2}$, in the $SU(2)$ doublet have the zero-mode wavefunctions,
\begin{eqnarray}
\lambda_{1/2} &: &
\Theta^{j,3}_{1/2}(z^1)\Theta^{1,1}_{1/2}(z^2)\Theta^{1,1}_{1/2}(z^3), 
\nonumber \\
\lambda_{-1/2} &: &
\Theta^{j,3}_{-1/2}(z^1)\Theta^{1,1}_{-1/2}(z^2)\Theta^{1,1}_{-1/2}(z^3), 
\end{eqnarray}
with $j=0,1,2$.
Here, these forms of wavefunctions are the same, i.e.
$\Theta^{j,3}_{1/2}(z^1) = \Theta^{j,3}_{-1/2}(z^1)$, 
$\Theta^{1,1}_{1/2}(z^2) = \Theta^{1,1}_{-1/2}(z^2)$ 
and $\Theta^{1,1}_{1/2}(z^3) = \Theta^{1,1}_{-1/2}(z^3)$, 
but we put the indices $\pm 1/2$ to show that they correspond 
to   $\lambda_{1/2}$ and 
$\lambda_{-1/2}$ components, respectively.
We embed the $Z_2$ twist $P$ in the $SU(2)$ gauge space as 
\begin{eqnarray}\label{eq:P-su2}
P = \left( 
\begin{array}{cc}
0 & 1 \\
1 & 0 
\end{array}
\right),
\end{eqnarray}
for the $SU(2)$ doublet.
In addition, we introduce a Wilson line along 
the Cartan direction of $SU(2)$, i.e.
\begin{eqnarray}
 \left( 
\begin{array}{cc}
1 & 0 \\
0 & -1 
\end{array}
\right),
\end{eqnarray}
e.g. on the first torus.
Since this Wilson line is not commutable 
with the orbifold twist $P$, 
the $SU(2)$ gauge group is completely broken.
Through this breaking, only three ${\bf 16}$ matter fields 
among three $({\bf 16},{\bf 2})$ fields
remain and their wavefunctions are obtained as 
\begin{eqnarray}
\Theta^{j,3}_{1/2}(z^1+C/2M_3)
\Theta^{1,1}_{1/2}(z^2)\Theta^{1,1}_{1/2}(z^3)
+
  \Theta^{3-j,3}_{-1/2}(z^1-C/2M_3) \Theta^{1,1}_{-1/2}(z^2) 
\Theta^{1,1}_{-1/2}(z^3), 
\end{eqnarray}
up to a normalization factor,
where $C$ denotes the continuous Wilson line 
on the first torus.
Similarly, we can introduce Wilson lines on other tori.
The orbifold twist $P$ acts trivially on 
the $({\bf 10},{\bf 1})_2$ fields, because 
they are singlets.
Then, six $({\bf 10},{\bf 1})_2$ fields among 
24 modes remain after orbifolding.

This model can realize the three families of ${\bf 16}$ matter fields 
as well as several (would-be Higgsino) fields ${\bf 10}$.\footnote{
They have no couplings such as ${\bf 16}{\bf 16}{\bf 10}$ 
(of the superpotential)
in bulk, because all of remaining matter fields have 
the same 6D chirality $(s_1,s_2,s_3)=(+,+,+)$ 
and their couplings are not allowed in the 10D Lagrangian.
Their couplings could be allowed on the orbifold fixed points, 
where the 10D Lorentz symmetry is violated.}
Furthermore, if we break $SO(10)$ to 
$SU(3) \times SU(2) \times U(1)_Y$ by Wilson lines 
on the third torus, 
the wavefunction profiles of ${\bf 16}$ matter fields 
would split such that the profiles of quarks and leptons have 
peaks at different points.

In the above example, we have embedded 
the non-trivial combination 
between orbifold twists and Wilson lines 
only into the $SU(2)$ part.
One can embed them into other parts 
and break e.g. $SO(10)$ by the orbifold twist.
Of course we can start with the $E_8$.
At any rate, non-trivial combinations of 
magnetic fluxes, Wilson lines and orbifold twists 
could lead to various interesting models.
Thus, it would be interesting to study more 
on this type of model constructions.

\section{Conclusion and discussion}

We have studied 10D super Yang-Mills theory with 
the gauge groups, $E_6$, $E_7$ and $E_8$.
On the torus and orbifold compactifications, 
we have considered the magnetic flux background 
as well as Wilson lines.
This type of model building leads to interesting models.
One of simple examples is the $E_6$ model on the torus 
and orbifold.
The $E_7$ and $E_8$ models would lead to various 
interesting models.
Thus, it would be interesting to investigate more 
on $E_7$ and $E_8$ models.
We would study them systematically elsewhere.

Also it would be interesting to study 
more on 4D effective theory, e.g. 
their flavor structure \cite{Abe:2009vi} and predictions on 
quark/lepton masses and mixing angles 
after we would construct 4D models with realistic spectra.
Although we have started with 10D theory, we can 
start with other extra dimensions, e.g. 
6D super Yang-Mills theory.

\subsection*{Acknowledgement}

K.-S.~C., T.~K., M.~M. and H.~O. are supported in part by the Grant-in-Aid for 
Scientific Research No.~20$\cdot$08326, No.~20540266, No.~21$\cdot$173 and
No.~21$\cdot$897 from the 
Ministry of Education, Culture, Sports, Science and Technology of Japan.
T.~K. is also supported in part by the Grant-in-Aid for the Global COE 
Program "The Next Generation of Physics, Spun from Universality and 
Emergence" from the Ministry of Education, Culture,Sports, Science and 
Technology of Japan.

\end{document}